\documentclass[11pt,onecolumn,fleqn]{article} %
\usepackage[fleqn]{amsmath} 
\usepackage{amsthm} 
\usepackage{amssymb}	
\usepackage[pdftex]{graphicx} 
\usepackage{multicol} 
\usepackage[dvips,letterpaper,margin=0.75in,bottom=0.75in]{geometry}
\usepackage[latin1]{inputenc}
\usepackage{amsfonts}
\usepackage{appendix}
\usepackage{cancel}
\usepackage[font=small,labelfont=bf]{caption}
\usepackage{fancyvrb}
\usepackage{gensymb}
\usepackage{verbatim}
\usepackage{wrapfig}
\usepackage{units}
\usepackage{url}
\usepackage{xfrac}
\usepackage{xcolor}

\makeatletter 
\renewcommand{\maketitle} 
{ \begingroup \vskip 10pt \begin{center} \large {\bf \@title}
	\vskip 10pt \large \@author \hskip 20pt \@date \end{center}
  \vskip 10pt \endgroup \setcounter{footnote}{0} }
\makeatother 
 
\newcommand{\abs}[1]{\left| #1 \right|} 
\newcommand{\pd}[2]{\frac{\partial #1}{\partial #2}} 
 
\let\baraccent=\= 
\renewcommand{\=}[1]{\stackrel{#1}{=}} 

\theoremstyle{definition}

\theoremstyle{remark}

\begin{document}
\title{\textbf{On the validity of drift-reduced fluid models for tokamak plasma simulation}}
\date{}
\author{Jarrod Leddy$^{1,2}$, Ben Dudson$^1$, and Michele Romanelli$^2$  \\
{\small E-mail: jbl504@york.ac.uk \\
$^1$York Plasma Institute, Department of Physics, University of York, Heslington YO10 5DD, UK \\
$^2$CCFE, Culham Science Centre, Abingdon, Oxfordshire OX14 3DB, UK}}

    \maketitle
    \begin{abstract}
    Drift-reduced plasma fluid models are commonly used in plasma physics for analytics and simulations; however, the validity of such models must be verified for the regions of parameter space in which tokamak plasmas exist.  By looking at the linear behaviour of drift-reduced and full-velocity models one can determine that the physics lost through the simplification that the drift-reduction provides is important in the core region of the tokamak.  It is more acceptable for the edge-region but one must determine specifically for a given simulation if such a model is appropriate.
    \end{abstract}

\section{Introduction}
Fluid models have been used to describe plasma behaviour in a magnetic field since Braginskii derived the plasma fluid equations and calculated the transport coefficients in 1965 \cite{Braginskii1965}.  Many other fluid systems have been derived based on these original equations using various simplifications to describe any specific physics of interest.  One such simplification that is often used is called the drift-reduction, in which the momentum equation is reduced by taking its curl resulting in an equation for the evolution of vorticity, $\vec{W}=\vec{\nabla}\times\vec{v}$.  For this new system to be closed, an assumption is made that the perpendicular velocities are dominated by the $\vec{E}\times\vec{B}$ drift, which relates the parallel vorticity to the potential: $W_{\parallel}=\nabla_{\perp}^2\phi$ (cgs Gaussian units are used for the duration of the paper).\\
\\
When this technique is used, time-scales below the ion cyclotron time are averaged over so fast waves are removed from the system.  Also, the effect of the pressure gradient on the velocity evolution is not carried into the vorticity equation as the curl of a gradient is zero.  This is especially significant because the largest pressure gradients will be perpendicular to the magnetic field and it is the perpendicular velocity equations which are replaced by an equation for the vorticity.\\
\\
These dropped physical effects will only be important in particular regions of parameter space, so this reduction is valid only when the missing physics is negligible.  By using linearisation techniques, the behaviour of these models can be compared to determine in which cases the drift-reduction is acceptable.  The systems will be simplified to the incompressible limit (such that $\vec{\nabla}\cdot\vec{v}=0$) to look at the most basic case that will still produce drift-waves.  Any differences for this case, then, are fundamental and will carry on into more complex scenarios.  Tokamaks are operated in well-defined yet broad parameter spaces, so the application of drift-reduced plasma fluid models for tokamak modelling can to be explicitly explored.
\section{Full velocity vs drift-reduced models}
A full-velocity model is one that evolves all three components of equation \ref{eqn:fullv_momentum}, the ion momentum equation \cite{Knight2012},\\
\begin{equation}
\begin{aligned}
mn\left(\pd{\vec{v}}{t}\right. &+ \left.\vec{W}\times\vec{v}\right) = \frac{\vec{J}\times\vec{B}}{c}-\nabla p  \\ 
&-\frac{mn}{2}\nabla(\vec{v}\cdot \vec{v})- mn\chi_v(\vec{\nabla}\times{\vec{W}}) \label{eqn:fullv_momentum}
\end{aligned}
\end{equation}
where $m$ is the mass of the ions, $n$ is the ion density, $\vec{v}$ is the ion velocity, $\vec{W}$ is the ion vorticity, $\vec{J}$ is the current density, $\vec{B}$ is the magnetic field, $p$ is the total pressure, $\chi_v$ is the velocity diffusivity, and $c$ is the speed of light.  By taking the curl of equation \ref{eqn:fullv_momentum}, an equation for vorticity is obtained.  It is convenient to take the parallel component of the vorticity equation, as shown in equation \ref{eqn:parallel_curl}, because it includes the behaviour of the perpendicular velocities:
\begin{equation}
W_{\parallel} = \hat{b}\cdot\left(\vec{\nabla}\times\vec{v}\right) = \left( \pd{v_x}{z} - \pd{v_z}{x} \right) \hat{y} \label{eqn:parallel_curl}
\end{equation}
where $\hat{x}$ and $\hat{z}$ are the perpendicular directions and $\hat{y}$ is parallel to the magnetic field line.  This geometry will be explained in more detail in the next section.  The normalised drift reduced equations, as derived by Hazeltine, et al. \cite{Hazeltine1985}, can then be written in the incompressible limit as follows:
\begin{equation}
\begin{aligned}
\pd{p}{t} &= -\left[ \phi, p \right] \\
\pd{W_{\parallel}}{t} &= -\left[ \phi, W_{\parallel} \right] - \nabla_{\parallel}J_{\parallel} \\
\pd{A_\parallel}{t} &= -\nabla_{\parallel}\phi + \eta J_{\parallel} + \nabla_{\parallel} p
\end{aligned} \label{eqn:drift-reduced system}
\end{equation}
where $p$ is the pressure, $\phi$ is the electric potential, $A_\parallel$ is the parallel vector potential, $\left[f,g\right]=\pd{f}{x}\pd{g}{z} - \pd{f}{z}\pd{g}{x}$ are the standard advection brackets, $J_{\parallel} = \nabla_{\perp}^2A_{\parallel}$, $W_{\parallel} = \nabla_{\perp}^2\phi$, and $\eta$ is the parallel resistivity.  The equation for $v_{\parallel}$ does not couple to these in the incompressible limit so is omitted, but it is important to note that the parallel velocity will evolve to maintain $\vec{\nabla}\cdot \vec{v}=0$.\\
\\
Using the same normalisations, the full-velocity model in the incompressible limit is given by
\begin{equation}
\begin{aligned}
\pd{p}{t} &= \nabla p \cdot \vec{v} \\
\pd{v_x}{t} &= \left(\vec{J}\times\vec{B}\right)_x - \nabla_x p \\
\pd{v_z}{t} &= \left(\vec{J}\times\vec{B}\right)_z - \nabla_z p \\
\pd{A_{\parallel}}{t} &=  - \nabla_{\parallel}\phi + \eta J_{\parallel} + \nabla_{\parallel} p \\
\pd{A_z}{t} &= 0 \;\;=\;\;  - \nabla_z\phi + \eta_{\perp} J_z + \nabla_z p \\
   &+ \left(\vec{v}\times\vec{B}\right)_z + \left(\vec{J}\times\vec{B}\right)_z
\end{aligned}
\end{equation}
with $\vec{\nabla}\cdot\vec{J}=0$, $\vec{J} = \vec{\nabla}\times\vec{B}$, and $\vec{B} = \vec{\nabla} \times \vec{A}$.  To evolve the vector potential, the generalised Ohm's law given by Lifshitz \cite{Lifshitz1981} is used with temperature gradients neglected due to the isothermal assumptions.  The equation for $A_x$ is excluded above because it simply evolves to maintain force balance without coupling to the remaining equations.  For both systems parallel derivatives are taken along the perturbed field by defining $\nabla_{\parallel}f = \partial_{\parallel}f - \left[ A_{\parallel},f \right]$.\\
\\
It is important when exploring the effects of the drift-reduction to ensure the two systems (full-velocity and drift-reduced) are identical in all other aspects.  To do this, the full-velocity system was drift-reduced and in the linear limit reproduces exactly the dispersion relation of the Hazeltine model, which can be seen in the next section in equation \ref{eqn:disp_rel_dr}.  In this way, the effects of the gyro-viscous cancellation which is used in both models are not observed in our comparison.

\begin{figure}[h]
\centerline{\includegraphics[width=0.4\linewidth]{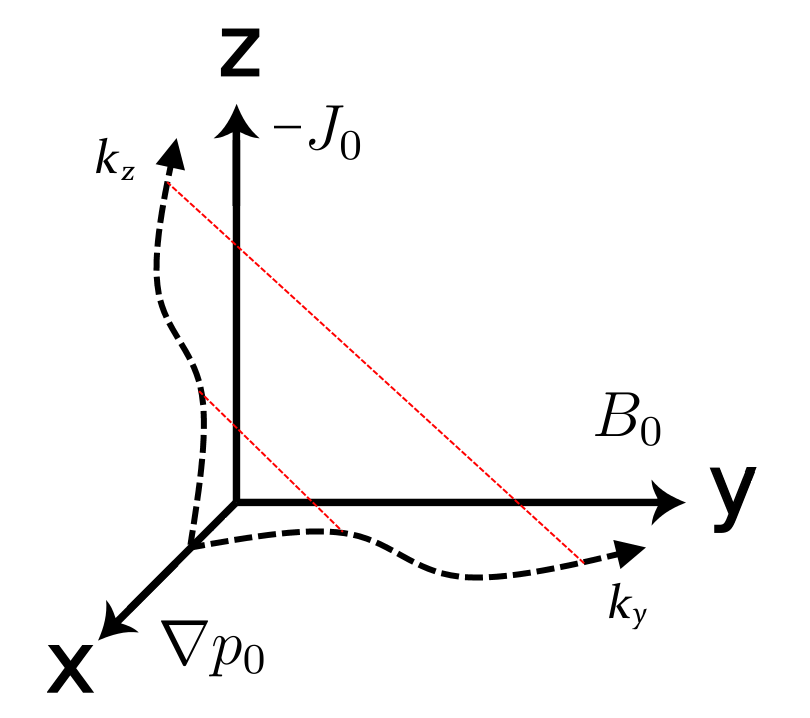}}
\caption{The geometry for the linearisation is quasi-3D with equilibrium pressure gradient, current density, and magnetic field that satisfy force balance.  Perturbations are in $y$ and $z$ such that the total perturbation is at an angle to the magnetic field, $B_0$.}
\label{fig:linearisation_coords}
\end{figure}
\section{Linearisation}
For all linearisations, we define a quasi-3D, orthogonal coordinate system ($x$-$y$-$z$) such that the equilibrium magnetic field $B_0$ is in the $y$-direction, the equilibrium current density $J_0$ is in the negative $z$-direction, and the background pressure gradient is in the $x$-direction; however, perturbations are only in $y$ and $z$ with no extent in $x$ consistent with a local approach, as detailed in figure \ref{fig:linearisation_coords}.  \\
\\
The background pressure gradient is present to drive the drift-wave instability, and the background current density and magnetic field are provided to satisfy force balance.   All perturbations are of the form $\tilde{f}=e ^ {i\left(k_y y + k_z z - \Omega t\right)}$ where $\Omega$ is the complex frequency defined as $\Omega=\omega + i\gamma$.  Both systems were un-normalised prior to linearisation so that physical parameters, such as Alv\'en speed, could be more easily substituted into the resulting dispersion relations.
\subsection{Drift-reduced dispersion relation}
The drift-reduced system in equation \ref{eqn:drift-reduced system}, once linearised, results in the following dispersion relation:
\begin{equation}
\begin{aligned}
\Omega^3 +& \left( \omega_* + i\eta\frac{v_A^2k_z^2\omega_{pi}^2}{4\pi\omega_{ci}^2} \right)\Omega^2 \\ 
&- \left(v_A^2k_y^2\right)\Omega - \left( v_A^2k_y^2\omega_* \right) = 0 \label{eqn:disp_rel_dr}
\end{aligned}
\end{equation}
where $\omega_*$ is the drift-wave frequency, $\omega_{ci}$ is the ion cyclotron frequency, $\omega_{pi}$ is the ion plasma frequency, and $v_A$ is the Alfv\'en speed.  These are defined as
\begin{eqnarray*}
&\omega_* = \frac{\nabla p_0 k_z}{m_in_0\omega_{ci}} \;\;\;\;\;\;\;\;\;\;\;\;\;\;\;\; &\omega_{ci} =  \frac{eB_0}{m_ic} \\
&\omega_{pi} = \sqrt{\frac{4\pi n_0e^2}{m_i}} \;\;\;\;\;\;\;\;\;\;\;\;\; 
&v_A  = \frac{B_0^2}{\sqrt{4\pi m_i n_0}}.
\end{eqnarray*}
Parallel Alfv\'en waves as well as resistive drift-waves can be seen in the terms of the dispersion relation.  In the case of zero resistivity the waves are stable and simply propagate.  
For $\eta>0$ the most unstable growth rate and corresponding frequency can be extracted using typical values for magnetic field, pressure gradient, and background density within a tokamak ($B=$1T, $\nabla p=$10$^6$Pa/m, $n_0=$10$^{18}$m$^{-3}$, and $k_y=k_z=1$).  
\begin{figure}[t]
\centerline{\includegraphics[width=0.5\linewidth]{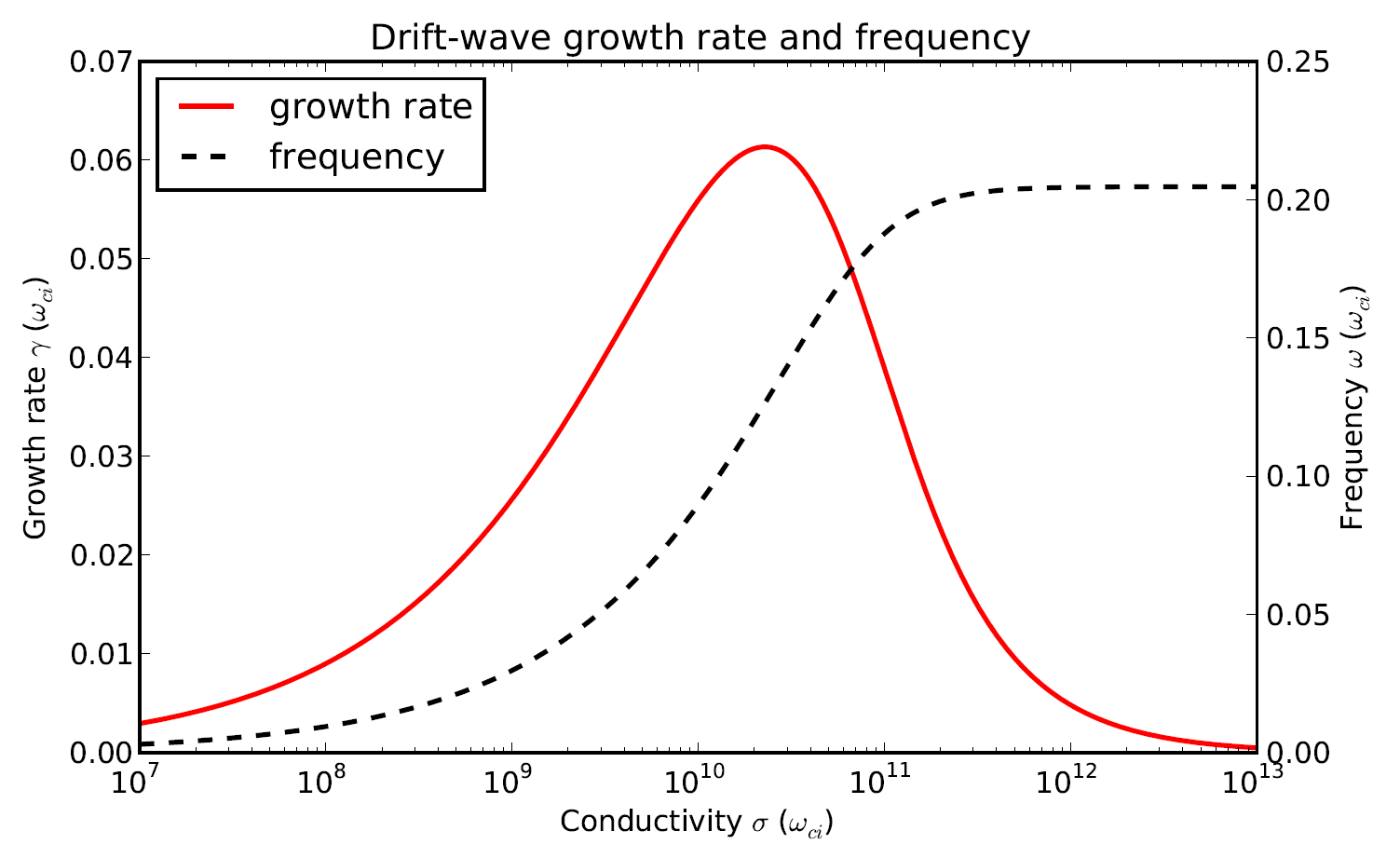}}
\caption{Growth rate (dashed) and frequency (solid) of the drift-wave instability as a function of conductivity for the drift-reduced system.  Frequency, growth rate, and conductivity are all normalised to the ion cyclotron frequency.}
\label{fig:DR_growthANDfrequency}
\end{figure}
These values are also chosen to satisfy $\omega_* < \omega_{ci}$ such that the ion cyclotron frequency is the highest frequency in the system (note that the pressure gradient here is typical for the pedestal and will be lower in other areas of the tokamak, reinforcing this ordering).  
The resulting frequency and growth rate are plotted in figure \ref{fig:DR_growthANDfrequency} as a function of conductivity, $\sigma = 1/\eta$ (ie. the inverse of resistivity).\\
\subsection{Full-velocity dispersion relation}
When the full-velocity system is linearised, the resulting dispersion relation (equation \ref{eqn:disp_rel_FV}) contains the extra, fast-physics that was lost in the drift-reduction.
\begin{equation}
\begin{aligned}
\left( \frac{\omega_{pi}^2\eta}{2\pi i \omega_{ci}^2} \right)\Omega^4 &+ \left( 1 +\frac{v_A^2k_y^2}{\omega_{ci}^2} + \frac{\omega_{pi}^4v_A^2 \left(k_y^2+2k_z^2\right)\eta^2}{16\pi^2\omega_{ci}^4} + \frac{\omega_{pi}^2\omega_*\eta}{\pi i\omega_{ci}^2} \right)\Omega^3  \\
&+ \left( \omega_* + i\eta\frac{v_A^2 \left(4k_y^2 + k_z^2\right)\omega_{pi}^2}{4\pi\omega_{ci}^2} \right)\Omega^2 - \left(v_A^2k_y^2 \right)\Omega - \left( v_A^2k_y^2\omega_*\right) = 0 \label{eqn:disp_rel_FV}
\end{aligned}
\end{equation}
\noindent The perpendicular resistivity has been approximated to be  double the parallel resistivity, as given in Wesson \cite{Wesson2011}.  Notice this expression has a higher order in $\Omega$ compared with equation \ref{eqn:disp_rel_dr} due to the additional equation for the perpendicular velocities, which results in an extra mode in the growth rate and frequency.  The two dispersion relations, equations \ref{eqn:disp_rel_dr} and \ref{eqn:disp_rel_FV}, are identical when $\Omega^4$ term, the last three terms in $\Omega^3$, and the parallel wave number in the second term of the $\Omega^2$ are neglected, indicating that these terms contain the physics lost in the drift-reduction.  This includes various propagating parallel and perpendicular resistive modes and the ion cyclotron wave. \\
 \begin{figure*}[t]
 \centerline{\includegraphics[width=0.9\linewidth]{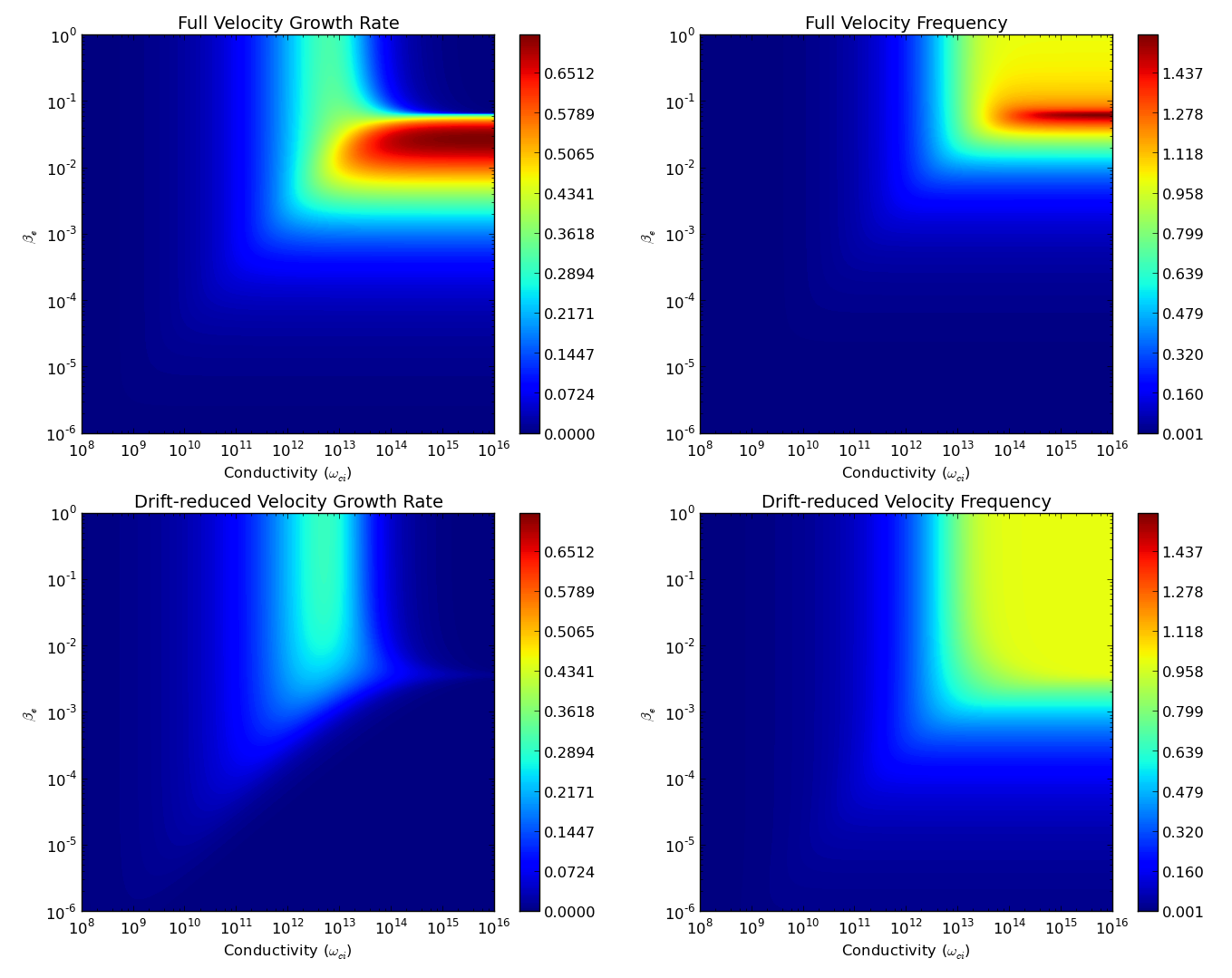}}
 \caption{Full velocity and drift-reduced growth rates and frequencies as a function of conductivity and $\beta_e$ at $B_0=1$T.  Growth rates and frequencies are normalised to the drift-wave frequency, $\omega_*$.}
 \label{fig:dr_and_fr_contour}
 \end{figure*}
\\
For the same values of magnetic field, background density, and pressure gradient the growth rate and frequency of this expression are quite similar to that of the drift-reduced model, so it was necessary to look at the solutions over a large parameter space in conductivity and electron beta, defined by
\begin{equation*}
\beta_e = \frac{p_{gas}}{p_{mag}} = \frac{8\pi n T_e}{B^2}.
\end{equation*}
In figure \ref{fig:dr_and_fr_contour} the magnetic field is set to constant $B=1$T and the density is adjusted to vary beta.  This is useful to do because the terms in equation \ref{eqn:disp_rel_FV} are not functions of only $\beta_e$ - they depend on various combinations of density and magnetic field.  In essence, the parameter space is 3D, however this is not easily visualised so magnetic field has been held constant for illustrative purposes.
\section{Tokamak relevance}
\begin{figure}[h]
\centerline{\includegraphics[width=0.5\linewidth]{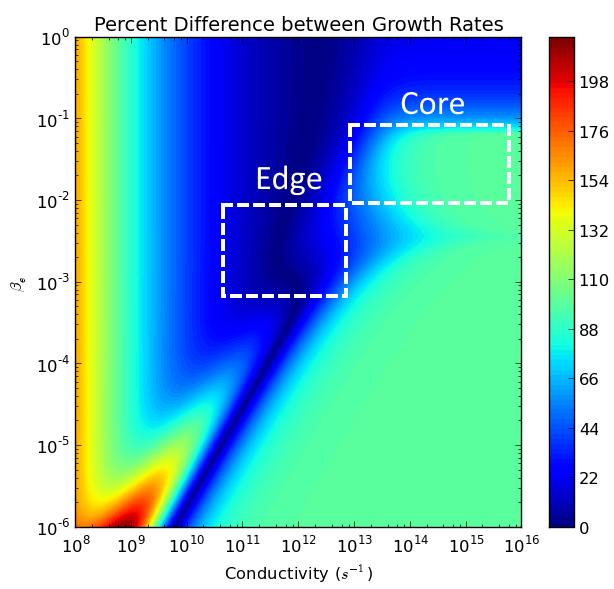}}
\caption{The percent difference from equation \ref{eqn:perc_diff} as a function of conductivity and electron beta.  Usual operational regimes for the core and edge in JET are marked.}
\label{fig:tokamak_parameters_space}
\end{figure}

The parameter space in which tokamaks operate is specific to the region within the tokamak (core vs edge) and the particular tokamak in question.  For a large tokamak of size similar to JET, the Joint European Torus at the Culham Science Centre, the core operates around $\beta_e=0.03$ and $\sigma=10^{15}$s$^{-1}$, while in the edge $\beta_e=0.005$ and $\sigma=10^{12}$s$^{-1}$.\\
\\
Figure \ref{fig:tokamak_parameters_space} depicts the percent difference in growth rates between the full-velocity and drift-reduced systems as given by
\begin{equation}
\Delta_{\%} = \abs{\frac{\gamma_{FV} - \gamma_{DR}}{\gamma_{FV}}}. \label{eqn:perc_diff}
\end{equation}
It is clear that at low conductivity the drift-reduction breaks down for all values of $\beta_e$.  This is due to the terms exclusively in equation \ref{eqn:disp_rel_FV} that are functions of $\eta$ and $\eta^2$ becoming very large at low conductivity, $\sigma = \eta^{-1}$.\\
\\
At low $\beta_e$, which corresponds to low density, we see a fairly universal disagreement between the models.  Since $\omega_* \propto n^{-1}$ and $v_A\propto n^{-1/2}$ these plasma parameters become larger at low density.  The plasma frequency $\omega_{pi}\propto n^{1/2}$, so it becomes small at low density, while the ion cyclotron frequency is not a function of density.  All of the additional terms in equation \ref{eqn:disp_rel_FV} vanish at low density except for the second term of the $\Omega^3$ term, $\frac{v_A^2k_y^2}{\omega_{ci}^2}$ which is proportional to $n^{-1}$.  At low density and high conductivity this term dominates, but as conductivity is lowered, the $\eta^2$ term takes over, thus the small area of agreement even at low $\beta_e$.\\
\\
The indication is that drift-reduced models are able to accurately reproduce edge behaviour, where errors can be as low as 0\%.  There are regions even in the edge, however, where a drift-reduced model may not be appropriate and errors can reach as high as 100\%.  In the core, there is a fairly consistent error of around 100\% from the full-velocity solution, so full-velocity fluid or gyro-kinetic models should be used in this region.
\section{Conclusion}
Drift-reduced models provide simplified dispersion relations for more succinct analytics and the exclusion of fast waves allows for larger timesteps leading to faster simulations, so these models are an important subset of the full fluid description.  The validity of these models has been tested for a simple quasi-3D slab resulting in drift-wave growth rates and frequencies that only agree with the full-velocity fluid description in specific regions of parameter space.  Though the worst agreement lies outside of the operational regime of tokamaks, there is still questionable agreement for core simulations.  It is important when choosing a model to use for tokamak plasma simulations to identify the parameter space in which the simulation will be operating as to identify whether a drift-reduced model is appropriate or if a more accurate, full-velocity model should be used instead.\\
\subsubsection*{Acknowledgements}
We thank Sarah Newton and Anantanarayanan Thyagaraja at CCFE for useful discussions guiding the evolution of this work. This work has received funding from the RCUK Energy Programme [grant number EP/I501045]\\

\begin{thebibliography}{1}

\bibitem{Braginskii1965}
S.~Braginskii.
\newblock {Transport Processes in a Plasma}.
\newblock {\em Reviews of Plasma Physics}, 1:205--311, 1965.

\bibitem{Hazeltine1985}
R.~D. Hazeltine, M.~Kotschenreuther, and P.~J. Morrison.
\newblock {A four-field model for tokamak plasma dynamics}.
\newblock {\em Physics of Fluids}, 28(8):2466, 1985.

\bibitem{Knight2012}
P.~Knight, A.~Thyagaraja, T.~Edwards, J.~Hein, M.~Romanelli, and K.~McClements.
\newblock {CENTORI: A global toroidal electromagnetic two-fluid plasma
  turbulence code}.
\newblock {\em Computer Physics Communications}, 183(11):2346--2363, Nov. 2012.

\bibitem{Lifshitz1981}
E.~M. Lifshitz and L.~P. Pitaevskii.
\newblock {\em {Course of Theoretical Physics: Physical Kinetics}}, volume~10.
\newblock 1981.

\bibitem{Wesson2011}
J.~Wesson and D.~Campbell.
\newblock {\em Tokamaks}.
\newblock International Series of Monographs on Physics. OUP Oxford, 2011.

\end{thebibliography}

\end{document}